\documentclass[12pt]{article}

\pdfoutput=1

\usepackage{bm,amsmath,amssymb,latexsym,graphicx,euscript,multirow,color,url,verbatim}
\usepackage{epsf}
\usepackage[compress,numbers]{natbib}
\usepackage{enumerate}
\usepackage[colorlinks=true,urlcolor=blue,anchorcolor=blue,citecolor=blue,filecolor=blue,linkcolor=blue,menucolor=blue,pagecolor=blue]{hyperref}

\allowdisplaybreaks

\makeatletter
\renewcommand{\@dotsep}{1000}
\makeatother

\newcommand{\captionfonts}{\small}
\makeatletter
\long\def\@makecaption#1#2{%
  \vskip\abovecaptionskip
  \sbox\@tempboxa{{\captionfonts #1: #2}}%
  \ifdim \wd\@tempboxa >\hsize
    {\captionfonts #1: #2\par}
  \else
    \hbox to\hsize{\hfil\box\@tempboxa\hfil}%
  \fi
  \vskip\belowcaptionskip}
\makeatother

\newcommand{\Sla}[1]%
{\kern0.12em{\raise.15ex\hbox{$/$}\kern-.74em #1}}

\long\def\symbolfootnote[#1]#2{\begingroup%
\def\thefootnote{\fnsymbol{footnote}}\footnote[#1]{#2}\endgroup}

\newcommand{\be}{\begin{equation}}
\newcommand{\ee}{\end{equation}}

\newcommand{\bea}{\begin{eqnarray}}
\newcommand{\eea}{\end{eqnarray}}
\newcommand{\mat}{\begin{pmatrix}}
\newcommand{\rix}{\end{pmatrix}}

\renewcommand{\bar}{\overline}
\renewcommand{\slash}[1]{#1\!\!\!/}

\newcommand{\go}{{\tilde g}}

\newcommand{\Ho}{{\tilde H}}

\newcommand{\cho}{{\tilde \chi}}
\newcommand{\sq}{{\tilde q}}
\newcommand{\st}{{\tilde t}}
\newcommand{\sbo}{{\tilde b}}

\newcommand{\beqa}{\begin{eqnarray}}
\newcommand{\eeqa}{\end{eqnarray}}

\newcommand{\beq}{\begin{equation}}
\newcommand{\eeq}{\end{equation}}

\newcommand{\abs}[1]{\left\vert#1\right\vert}

\newcommand{\order}[1]{{\cal O}\left(#1\right)}

\newcommand{\met}{{\slash E_T}}

\definecolor{white}{rgb}{1.0,1.0,1.0}
\newcommand{\enwhiten}[1]{{\color{white} #1}}

\begin{document}

\begin{titlepage}

\begin{flushright}
\small{RUNHETC-2014-01}\\
\end{flushright}

\vspace{0.5cm}
\begin{center}
\Large\bf
A Swarm of $B$s
\end{center}

\vspace{0.2cm}
\begin{center}
{\sc Jared A. Evans\symbolfootnote[4]{jaevans@physics.rutgers.edu}
}\\
\vspace{0.6cm}
\small \textit{New High Energy Theory Center\\
Rutgers University, Piscataway, NJ 08854, USA}
\end{center}

\vspace{0.5cm}
\begin{abstract}
\vspace{0.2cm}
\noindent

New physics signals containing five or more $b$-tagged jets, but without $\met$ or leptons, could realistically be sitting within the current 8 TeV LHC data set without receiving meaningful constraints from any of the existing LHC searches at either ATLAS or CMS.  This work provides several examples of simple, motivated models that yield final states containing many $b$-jets.  To study the potential for uncovering new physics in these high $b$-jet multiplicity channels, this paper focuses on a natural supersymmetry scenario where each of the pair-produced stops decays to an on-shell chargino, which subsequently decays via an MFV-motivated, $R$-parity violating coupling.  This gives rise to an eight-jet final state containing six $b$-quarks.  Although no public measurements exist, estimates indicate that the standard model backgrounds in high $b$-jet multiplicity channels should be very small.  To circumvent the background uncertainty, an asymmetric method is presented that utilizes two different techniques to conservatively exclude or to discover new physics in high $b$-jet multiplicity final states.

\end{abstract}
\vfil

\end{titlepage}

\tableofcontents

\section{Introduction}
\label{sec:intro}

Run I of the LHC at both 7 and 8 TeV has been a great success.  A Higgs-like particle has been discovered \cite{Chatrchyan:2012ufa,Aad:2012tfa}.  Many models are now heavily constrained or even excluded.  Among these constrained models is supersymmetry (SUSY), which is widely regarded as one of the most plausible explanations  for the origin of the electroweak scale (for recent reviews and references, see~\cite{Craig:2013cxa,Feng:2013pwa}).  
The absence of LHC signatures indicative of simple low-energy SUSY spectra, typically involving large missing energy ($\met$), is at odds with SUSY providing a satisfactory resolution the electroweak hierarchy problem.   
  While this lack of substantial deviations from the standard model in LHC searches has been disappointing, it strongly motivates an initiative to leave no signature uncovered  -- particularly those arising from more complicated SUSY models with non-generic spectra. 

All searches for new physics require some discriminant to distinguish a signal from the standard model backgrounds.  Many searches for new physics utilize simple requirements on variables such as $\met$, $H_T$, or number of leptons.  Others employ significantly more complicated discriminants, for instance $\alpha_T$ or $M_R$ \cite{Randall:2008rw, Rogan:2010kb}.    Among the existing LHC searches, requiring one or more $b$-tagged jets can be very useful in removing certain standard model backgrounds.   In this paper, a search is proposed that is centered around using a very high $b$-tagged jet multiplicity ($\geq 5$) as the main discriminant for distinguishing signal from background.  This particular discriminant has been used before in a lepton-based study \cite{CMS:2013qua}, but never in an all-hadronic study.

Many well-motivated models of new physics can give rise to a high $b$-jet multiplicity signature.  For example, supersymmetric models with violated $R$-parity or hidden valley extensions (such as stealth SUSY models) can have many $b$-jets, while exhibiting insufficient missing energy to fall under the more traditional SUSY search strategies.   Alternatively, fourth-generation quarks or the cascade decays between resonances of extended Higgs sectors (for instance, of the type IV 2HDM~\cite{Craig:2012vn}) can produce high $b$-jet multiplicity final states. 

However, the many $b$-jet background of the standard model (a subset of the QCD background) is very uncertain.  To date, there has been no public LHC measurement of many $b$-tagged jets without accompanying $\met$ or leptons.   As we will discuss, both monte carlo generation and simple projections estimate that the background is small.  However, very large NLO $K$-factors have been known to appear within QCD backgrounds \cite{Rubin:2010xp}.    It is extremely difficult to estimate how large the backgrounds are expected to be without using data-driven methods.  For this reason, we will utilize a novel asymmetric approach for treating signal exclusion and signal discovery separately.  

This paper is outlined as follows:  in Section~\ref{sec:NP}, we will briefly discuss some new physics models that give rise to high $b$-jet multiplicities.  
 We will then propose an LHC study looking for new physics in high $b$-multiplicity channels in Section~\ref{sec:btags}.  First, we estimate the highly uncertain QCD backgrounds in subsection~\ref{sec:bkgs}, before presenting an asymmetric method to separately constrain (\ref{sec:sigEx}) and discover (\ref{sec:sigDisc}) new physics.  Section~\ref{sec:conclusions} contains the conclusions.

\section{New physics with high $b$-jet multiplicities}
\label{sec:NP}

Final states with many $b$-jets, but without leptons or large $\met$, do not only originate within baroque models.  In fact, ``third-generation dominance,'' a simple ansatz for the coupling of new particles to the standard model particles, naturally leads to decays that involve $b$-quarks.  One motivation for this ansatz is that any new physics closely tied to the generation of mass, such as extended Higgs sectors, is expected to exhibit this behavior.  Additionally, indirect bounds from low energy flavor and CP observables, such as $K\!-\!\bar K$ mixing and the neutron electric dipole moment, indicate that the standard model's first- and second-generation particles must couple to new physics either very weakly or in a very structured manner, while the bounds on interactions involving the third-generation are generically much weaker.   While this is more of a statement concerning current experimental limitations pertaining to flavor studies on the third-generation, the possibility that sizable couplings exist is enough motivation to search for models with such couplings.  Minimal flavor violation (MFV) \cite{D'Ambrosio:2002ex}, which ties all flavor relations to the standard model yukawas, is one example of third-generation dominance, however, more general scenarios of third-generation dominance are quite plausible \cite{Agashe:2005hk, Nomura:2007ap, Evans:2013kxa}. 

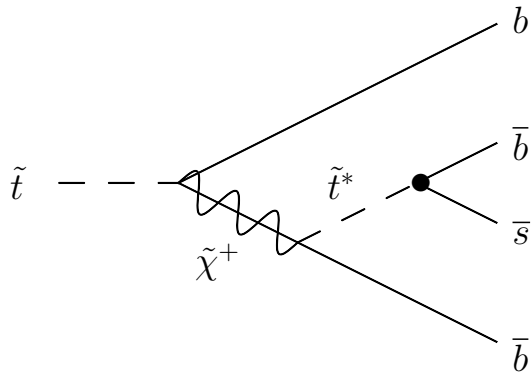
\begin{figure}[t]
\begin{center}
\includegraphics[scale=1.6]{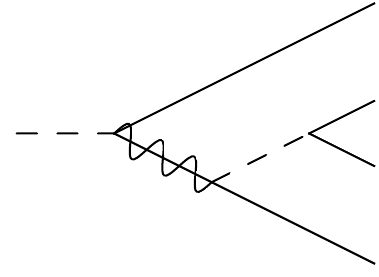}
\begin{picture}(0,0)(0,0)
\large
\put(-190,60){$\st$}
\put(0,123){$b$}
\put(0,75){$\bar b$}
\put(0,43){$\bar s$}
\put(0,-5){$\bar b$}
\put(-120,32){$\cho^+$}
\put(-70,60){$\st^*$}
\put(-39,62){\Large $\bullet$}
\normalsize
\end{picture}
\begin{minipage}[t]{7.0in}%
\caption{\label{fig:4bstop} Four-body $R$-parity violating decay of the stop.  The stop, $\st$, decays via the top yukawa into a $b$-quark and the intermediate, predominantly higgsino, on-shell chargino, $\cho^+$.  The chargino then decays back through the off-shell $\st$, which interacts via the $\lambda''_{323}$ RPV operator.  The net result is $\st \to b (\bar b \bar b \bar s)$.  Together with the anti-stop decay, this is nominally an eight-jet final state with six $b$-quarks. }
\end{minipage}
\end{center}
\end{figure} 

Arguably, the most motivated spectra from natural SUSY contain light stops and higgsinos \cite{Papucci:2011wy} (naturalness constraints on the gluino mass can be alleviated if the gauginos are Dirac \cite{Fox:2002bu,Kribs:2012gx}).   For $R$-parity violating couplings originating from an MFV structure \cite{Nikolidakis:2007fc,Csaki:2011ge,Krnjaic:2012aj,Franceschini:2013ne,Csaki:2013we}, the superpotential coupling $\lambda''_{323} U_3^c D_2^c D_3^c$ is the largest.  With its fairly large cross-section, pair-production of stops (i.e., through $R$-parity conserving diagrams) is a promising avenue for discovery of SUSY.  If a stop that is at least partially right-handed is the lightest superpartner, then the decay $\st \to \bar b \bar s$ would be the dominant decay channel in these MFV RPV SUSY models \cite{Franceschini:2012za,Bai:2013xla}.  However, if the higgsinos (which can naturally be nearly degenerate) are lighter than the stop, the decay $\st \to b \cho^+$ (and, when not kinematically forbidden, the phase-space suppressed $\st \to t \cho^0_{1,2}$) would dominate as this decay uses the top yukawa, an $\order{1}$ coupling.  The chargino can then decay promptly through an off-shell $\st$ to yield a six $b$-quark and two light quark final state \cite{Evans:2012bf,Evans:2013uwa} (see Fig.~\ref{fig:4bstop}),  
\beq
\st \to b \cho^\pm \to b (\bar b \st^*) \to b (\bar b \bar b \bar s).
\eeq
This branching ratio can realistically be near 100\%. 

In $R$-parity conserving SUSY, the addition of a hidden sector \cite{Strassler:2006im,Strassler:2008fv} could yield a very high multiplicity of $b$-jets without introducing significant $\met$.  Stealth SUSY \cite{Fan:2011yu,Fan:2012jf} is a simple example that can give rise to this signature.   If the stealth sector contains an NMSSM-like singlet and singlino (with $m_{\tilde S} \approx m_S$), then $S$ will most often decay to $b \bar b$ through its mixing with the Higgs sector.  In the presence of a jet $p_T$ hierarchy~\cite{Evans:2013jna}, gluinos as light as 800 GeV could be hiding in the data.  A simple model involves 
\beq
\go \to j \sq \to j j \Ho^0_1 \to jj S \tilde S \to jj SS \tilde G \to jj (b \bar b) (b \bar b),
\eeq
where $\sq$ is a second-generation squark, $m_\go\gg m_\sq$ introduces a jet $p_T$ hierarchy, and a singlet-singlino mass squeezing makes the gravitino too soft to contribute appreciably to the $\met$.  In principle, this is a signal with twelve jets containing eight $b$s.  Even with a jet $p_T$ hierarchy, this simple model may be constrained by searches sensitive to the rarer $S\to \tau^+\tau^-$ decays.  However, with decoupled gluinos (which is consistent with naturalness for Dirac gluinos), the direct pair production of higgsinos in a similar model,
\beq
pp \to \Ho^+  \Ho^0_i \to \Ho^0_1 \Ho^0_1 + \{\mbox{soft}\} \to S S \tilde S \tilde S \to SSSS \tilde G \tilde G \to (b \bar b) (b \bar b) (b \bar b) (b \bar b), 
\eeq
gives rise to eight jets, all of which are $b$-quarks.  As an alternative, one could have direct sbottom production, where the sbottom decays as,
\beq
\sbo \to b \tilde S \to b S   \tilde G \to b (b\bar b)
\eeq
giving rise to six jets, all of which are $b$-quarks.  One could argue that $S\to \tau^+\tau^-$ decays could constrain this model as well.  However, due to the paucity of $\tau+b$ searches at the LHC, there may be no sensitivity to this much lower $S_T$ signature \cite{Evans:2012bf,Evans:2013uwa}.  Even if such searches were to exist, after branching factors, soft partonic-level $\tau$s of $\order{30-40 \mbox{ GeV}}$ would only rarely produce events with useful light leptons.

Resonances of extended Higgs sectors can cascade decay into other resonances.  For instance, many $b$-jets can arise from production of a heavy Higgs that decays dominantly into two pseudoscalars, which each decays to a SM higgs and a light pseudoscalar, i.e.,
\beq
pp \to H \to AA \to \phi h \phi h \to (b \bar b) (b \bar b) (b \bar b) (b \bar b),
\eeq
where $A$ and $\phi$ are pseudoscalars, $H$ is the heavy Higgs, and $h$ is the observed $\sim125$ GeV SM-like Higgs state.  Although constrained by the current best fits to the observed Higgs state \cite{Craig:2013hca}, if the model is an extension of a Type III or IV  two Higgs doublet model (also known as ``lepton-specific'' and ``flipped,'' respectively)~\cite{Craig:2012vn}, then $b\bar b$ pairs can dominate all pseudoscalar decays.  With the high branching fractions and low expected backgrounds, it is plausible that multi-$b$ could prove the most sensitive channel to these models.  However, it is also possible that other decay channels of the Higgs, such as $h\to WW^*$ or $h\to \gamma\gamma$, could prove the more promising discovery mode.  Decay chains in extended Higgs sectors, such as the example shown here, are a motivated extension to the standard model wherein many $b$-jets can occur.

The decay of $b'\to b h \to b (b\bar b)$ makes vector-like fourth-generation quarks another place where high multiplicity $b$-jet signatures could appear.  Minimal $b'$ models require BR$(b' \to b Z)\sim$ BR$(b'\to b h)$ \cite{Aguilar-Saavedra:2013qpa}; in these models, if BR$(b'\to b h)\sim50\%$, then BR$(b'\bar b'\to6b)\sim 13\%$.  However, in extended models, such as those having both $(b' \; q_{-4/3})_{L/R}$ and $b''_{L/R}$ fields  (in analogy to the $t'$ models of \cite{Azatov:2012rj}), the branching ratios of $b'$ have more freedom, and $b'\to bh$ can dominate.  Although these models can receive significant constraints from precision electroweak observables, most notably $Z\to b\bar b$, a complete model may realize a consistent solution.  As we are only positing this decay path as a phenomenological possibility, uncovering a specific model is tangential to the focus of this work.  

For the remainder of this paper, we will focus on the $6b + 2j$ signature of pair-produced stops in natural MFV RPV SUSY (Fig.~\ref{fig:4bstop}).  This signal has been shown to receive no meaningful constraint from any existing LHC study \cite{Evans:2012bf,Evans:2013uwa}.  To remove the possibility of top quarks entering in decays (which could receive constraints from existing searches), we will consider signal regions where the decay $\st\to t \cho^0$ is negligible, i.e.~$m_\st-m_\cho \lesssim 200$ GeV.  The branching fraction into the desired final state can be set to one without invoking any artificial squeezing of new states or contrived assumptions about coupling structures.  After the branching fraction is set to one, this simplified model is left with only two free parameters:  the mass of the stop and the mass of the chargino.  

We will briefly note that displaced decays, including two-, three-, and four-body scenarios, are a realistic option for most of the signatures mentioned above.  While there are certainly difficulties with $b$-jets originating from a point away from the primary interaction vertex, this possibility is important enough that such signals should be specifically addressed in studies of displaced new physics.

Estimating the sensitivity to the various other signatures mentioned here is beyond the scope of this work.  However, we stress that high $b$-jet multiplicities can easily arise from a variety of new physics models. 

\section{A many $b$-tags study}
\label{sec:btags}

\begin{table}[t]
\begin{center}
{\small
\begin{tabular}{|c||c|c|c|c|c|} \hline
\multicolumn{6}{|c|}{\bf Preselection Cuts}    \\ \hline
\multicolumn{6}{|c|}{$H_T(p_{T}>40; \abs{\eta}<2.5 )>750$ GeV} \\  
\multicolumn{6}{|c|}{No isolated leptons with $p_T>20$ GeV, $\abs{\eta}<2.4$, and $I_{rel}<0.15$} \\
 \hline \hline
{\bf Cuts} & {\bf Region 1} & {\bf Region 2} & {\bf Region 3} & {\bf Region 4} & {\bf Region 5}  \\ \hline
$H_T$ (GeV)  & 750  & 1000  & 1250  & 1500  & 1750 \\ \hline
$b_{eff} \; (\%)$  & $50$  &\multicolumn{2}{c|}{$60$} & \multicolumn{2}{c|}{$70$} \\
$c_{eff} \; (\%)$  & $4.0$  &\multicolumn{2}{c|}{$9.0$}   & \multicolumn{2}{c|}{$19$} \\
$j_{eff} \; (\%)$  & $0.07$   &\multicolumn{2}{c|}{$0.30$} &\multicolumn{2}{c|}{$1.5$} \\ \hline
$n_b$ & \multicolumn{5}{c|}{ $\geq 5 \; b$-tagged jets w/ $p_T>30$ GeV and $\abs{\eta}<2.5$} \\ \hline
\end{tabular}
\caption{The cuts used in this study.  $H_T$ is, in all cases, the sum over jets with $p_T>40$ GeV with $ \abs{\eta}<2.5$.  In addition to using higher $H_T$ cuts, the different signal regions also use different $b$-tagging working points.  The parameters $b_{eff}$, $c_{eff}$, and $j_{eff}$ are the percent of jets originating from a partonic-level $b$, $c$, or light parton ($guds$) that are tagged as a $b$-jet.  Realistic efficiencies are here taken from the CMS 7 TeV study of $b$-tagging with the CSV tagging algorithm~\cite{Chatrchyan:2012jua}.
\label{tab:cuts}}
}
\end{center}  
\end{table}

While signatures containing five or more $b$-jets with neither $\met$ nor isolated leptons may be motivated from a new physics perspective, the challenge to such a study is in the background estimation.  However, even with an unknown background, it can be possible to place meaningful bounds.  In fact, if the signal alone (i.e., considered in the presence of zero background) would be too large to account for the observed data, then the signal can be excluded.   In this section, we present a simple study where our RPV stop signature can be conservatively constrained independent of the estimated background.  We will then illustrate how an already established technique could ferret out the resonant structures in the events and ultimately conclude that an excess is due to a genuine signal, rather than due to a larger than expected background.  Such an asymmetric approach to exclusion and discovery has been used in experimental studies before (see, for instance, \cite{ATLAS:2013tma}), so this general strategy is not without precedence.

One immediate concern for signals with high jet activity, but without leptons or $\met$, is whether the events will pass the trigger requirements.  Fortunately, CMS has in place six-jet, eight-jet, and high-$H_T$ triggers that have no prescaling instituted.  By the end of the 8 TeV run, these stood at:  6j -- $p_{T,j} >45$ GeV; 8j -- $p_{T,j} >30$ GeV; and $H_T$ -- $\sum(\mbox{over jets with }  p_{T}>40)\, p_{T,j} > 750$ GeV \cite{BDahmes}.\footnote{Additionally, there is a parked data trigger requiring 4 jets with $p_T>50$ GeV.  It is possible that this trigger could be useful for data-driven background determination.}  In this study, the $H_T$ trigger proved the most powerful probe for most signals, so it will be the only trigger used throughout this study.\footnote{For low and intermediate stop masses, far below the exclusion limit, some other considered signal regions proved slightly more effective at constraining the signal.  For simplicity, we utilize only the $H_T$-based search regions with $\geq5$ $b$-tagged jets in this work.  For completeness, we will note that requiring $N_j\geq8,N_b\geq5$ for several different choices of $p_{T,j/b}>30,\,40,\,50,\,60$ GeV, as well as $N_j\geq8,N_b\geq6, p_{T,j/b}>40$ GeV, each occasionally served as the most effective search region, although the gain over the $H_T$-based search regions was typically small.}  

As the trigger efficiencies in our region of interest are unknown, we assume they are 100\%.  Unless the trigger efficiencies are very low, the results will not be extremely sensitive to these details, as our signal is typically an order of magnitude larger than our backgrounds in regions where the trigger efficiency turn-on could be an issue.  We note that the $n_b=0$ QCD data could be used to reliably model the $H_T$ trigger efficiency turn-on. 

In this study, we use five different signal regions with differing $H_T$ thresholds.  Additionally, we utilize three different $b$-tagging working points (from the CMS CSV tagging algorithm~\cite{Chatrchyan:2012jua}) for the search -- medium, tight, and very tight -- with approximate $b$-tagging efficiencies of 0.7, 0.6, and 0.5, respectively.  We note that the ``very tight'' working point is taken from their figures, but has not been used in any CMS experimental study thus far.  For simplicity, in this work, $b$-tagging efficiencies are treated as being constant in both $p_t$ and $\eta$.  Details of the five search regions are shown in Table~\ref{tab:cuts}.

\subsection{Backgrounds}
\label{sec:bkgs}

\begin{table}[t]
\begin{center}
{\small
\begin{tabular}{|c|c||c|c|c|c|c|} \hline
\multirow{2}{*}{\bf Backgrounds} & \multirow{2}{*}{\bf K-factor} & \multicolumn{5}{|c|}{\bf Number of events in 20 fb$^{-1}$ at 8 TeV} \\ \cline{3-7}
       & &  {\bf Region 1} & {\bf Region 2} & {\bf Region 3} & {\bf Region 4} & {\bf Region 5}  \\ \hline\hline
$b\bar bb\bar b+\{$jets$\}$          & 3 & $5.3$ & $4.3$ & $1.3$ & $1.4$ & $0.5$ \\
$b\bar bb\bar bb\bar b$               & 3 & $0.5$ & $0.2$ & $<0.1$ & $<0.1$ & $<0.1$ \\
$b\bar bb\bar bc\bar c$                & 3 & $0.1$ & $0.1$ & $<0.1$ & $<0.1$ & $<0.1$ \\
$t\bar tb\bar b$                              & -- & $0.9$ & $1.3$ & $0.5$ & $0.7$ & $0.3$ \\ \hline 
\multicolumn{2}{|c||}{Total}                    & $\bf 6.8$ & $\bf 5.9$ & $\bf 1.9$ & $\bf 2.1$& $\bf 0.7$ \\ \hline
\multicolumn{2}{|c||}{``Observed'' Events for Fig.~\ref{fig:ExclusionLimits}}                    & $\bf 7$ & $\bf 6$ & $\bf 2$ & $\bf 2$& $\bf 1$ \\ \hline\hline\hline 
\multicolumn{2}{|c||}{\bf Signal}  & \multicolumn{5}{|c|}{\bf Number of events in 20 fb$^{-1}$ at 8 TeV} \\ \hline
     $m_\st$ (GeV)  & $m_{\cho^\pm}$ (GeV) & {\bf Region 1} & {\bf Region 2} & {\bf Region 3} & {\bf Region 4} & {\bf Region 5} \\ \hline\hline 
   $150$&$100$ & $\bf 64$ & $18$ & $5.1$ & $3.9$ & $1.3$ \\
   $300$&$200$ & $\bf 110$ & $78$ & $22$ & $16$ & $5.1$ \\
   $500$&$350$ & $45$ & $\bf 50$ & $20$ & $15$ & $5.7$ \\
   $700$&$600$ & $6.0$ & $13$ & $\bf 8.1$ & $7.9$ & $3.4$ \\
   $800$&$600$ & $2.7$ & $6.1$ & $5.1$ & $\bf 6.0$ & $2.9$ \\
   $900$&$875$ & $0.2$ & $0.6$ & $0.6$ & $1.1$ & $\bf 0.7$   \\ \hline 
\end{tabular}
\caption{{\bf Above:}  The backgrounds and how they populate the signal regions used in this study.  Hard process generation is in either {\tt Alpgen} or {\tt Madgraph 5}, as discussed in the text of Section~\ref{sec:bkgs}, and parton showering is performed by {\tt PYTHIA 8}.  $t\bar tb\bar b$ is normalized to the measured cross-section of \cite{CMS:2013vui}, while all other backgrounds receive a $K$-factor $=3$.  We stress that these very uncertain backgrounds are not ultimately necessary for our proposed method of setting limits.  In projecting a potential exclusion for Figure~\ref{fig:ExclusionLimits}, we round our expected background to act as our ``observed'' values.     {\bf Below:}  Six of our signal benchmark points are presented (indicated with $m_\st$ and $m_{\cho^\pm}$), and we display how they are expected to populate each of our five signal regions.  The most sensitive channel to that signal is emboldened.  Production of the signals is in {\tt Madgraph 5}; showering is in {\tt PYTHIA 8}.
\label{tab:bkgds}}
}
\end{center}  
\end{table}

The dominant backgrounds to signals with $\geq 5 \; b$s are $t\bar tb \bar b$, $b\bar bb\bar bb\bar b$, $b\bar bb\bar bc\bar c$, and $b\bar bb\bar b +\{$light jets$\}$.  With the exception of $t\bar tb\bar b$, these have never been explicitly measured in published LHC data.  The QCD cross-sections for each of these processes have very large uncertainties.  As it is difficult to reliably estimate the backgrounds, we will implement an asymmetric approach that is insensitive to the background estimation by using different methods for placing limits (subsection~\ref{sec:sigEx}) and for making discoveries (subsection~\ref{sec:sigDisc}).  However, we will first estimate the sizes that might be expected for these standard model backgrounds.

CMS has measured $t\bar t b \bar b$ explicitly in 8 TeV data and found $\sigma(t\bar t b \bar b)=460\pm45(\mbox{stat})\pm120(\mbox{sys})$ fb under a $b$-jet $p_T$-cut of 20 GeV \cite{CMS:2013vui}.  As this measurement was made within the relatively clean, semi-leptonic top decay channel, the size of the all-hadronic channel relevant for this study can be reliably scaled.  For the signal regions used in our study, we adopt a $p_T$-cut of 30 GeV, however, we will conservatively apply the measured $p_T>20$ GeV cross-section to our sample.  To simulate this background, we generated $t\bar t b \bar b+ \{0,1,2\}$ jets with {\tt Alpgen} \cite{Mangano:2002ea} and showered with {\tt PYTHIA 8} \cite{Sjostrand:2007gs}.  {\tt Alpgen} is not equipped to match jets in the 4Q sample \cite{Mangano:2002ea}, however, we combine these three samples and normalize their sum to the measured $\sigma(t\bar t b \bar b)$ of \cite{CMS:2013vui}.  As a cross-check, $t\bar t b \bar b$ was also generated in {\tt MadGraph 5} \cite{Alwall:2011uj}.  The normalized distribution there is in good agreement with the {\tt Alpgen} sample.  In each of our five signal regions, $\lesssim 1$ event is expected (see Table~\ref{tab:bkgds}).

The cross-section for $b\bar bb\bar bb\bar b$ ($b\bar bb\bar bc\bar c$) is unknown.  Na\"ive generation in {\tt MadGraph 5} yields a partonic-level 8 TeV cross-section of 11 fb (35 fb) under the cuts $p_{T,j} \geq 30$ GeV and $\abs{\eta}<2.5$.  Of course, six jets with $p_T$ of 30 will not pass the trigger thresholds, and we found that typically no events between the two samples would be expected to pass the selection cuts of our signal regions.  For our presentation, we apply a $K$-factor of $K=3$ to that sample (resulting in a cross-section of 33 (105) fb).  This $K$-factor is chosen because it is greater than typical top system $K$-factors, and thus we believe it to be a conservative choice.  However, even with this $K$-factor, these two samples together yield fewer than one event in each signal region.

The cross-section for $b\bar bb\bar b+\{$light jets$\}$ is also unknown.  To simulate this, we use {\tt Alpgen} to  generate $b\bar b b \bar b+ \{0,1,2,3\}$ jets.  Again, {\tt Alpgen} is not equipped to match jets in the 4Q sample \cite{Mangano:2002ea}, however, we scale these four samples up by a K-factor of 3, yielding an overall cross-section of 88 pb.\footnote{{\tt Alpgen} technically includes the $b\bar bb\bar bb\bar b$ and $b\bar bb\bar bc\bar c$ in this sample \cite{Mangano:2002ea}.  However, as a direct generation of this is shown to be quite small, we ignore this double counting.}  As can be seen in Table~\ref{tab:bkgds}, this proves the dominant background in all signal regions.  However, there are very large uncertainties on this estimation.  It has been established that large $K$-factor corrections can sometimes appear in QCD backgrounds \cite{Rubin:2010xp}, especially along the tails of distributions.  For this reason, we are hesitant to utilize leading order monte carlo estimations in setting exclusion limits.\footnote{On the other hand, extremely good agreement has also been measured between some monte carlo distributions and LHC data \cite{CMS:2012nca}.}  Fortunately, as we will discuss in the next subsection, the method we utilize in this study is designed to remove all sensitivity to the background estimations.  

As an alternative to the extremely conservative exclusion strategy we will use in this work, it is possible to attempt to utilize data-driven techniques to estimate the backgrounds.  Unfortunately, projections from a control region with $n_b=5$ are nearly impossible, because the signal would be expected to heavily contaminate any such control region that contains events.  However, one could use projections from control regions  with $n_b=0-4$ to estimate the size of the $n_b=5$ backgrounds.  This would be particularly useful in measuring the dominant background of $b\bar bb\bar b+\{$light jets$\}$.  Similarly, while it is expected to be small, an ansatz of
\beq
\frac{\sigma (b\bar bb\bar bj j)}{\sigma (b\bar bb\bar bb \bar b)} \approx \frac{\sigma (t\bar tj  j)}{\sigma (t\bar tb \bar b)} \approx 2.3\%
\eeq
could prove useful in making a controlled extrapolating from $n_b=3,4$ measurements to get estimations for the $b\bar bb\bar bb\bar b$ and $b\bar bb\bar bc\bar c$ backgrounds.\footnote{The $t\bar t b \bar b$ and $t\bar t jj$ cross-section data are taken from \cite{CMS:2013vui}.}

\subsection{Signal exclusion}
\label{sec:sigEx}

\begin{figure}[t]
\begin{center}
\hspace{7mm} \large$m_{\st}$ vs.~$\delta \equiv m_{\st}-m_{\cho^\pm}$ \\
\hspace{7mm}
\includegraphics[scale=1.3]{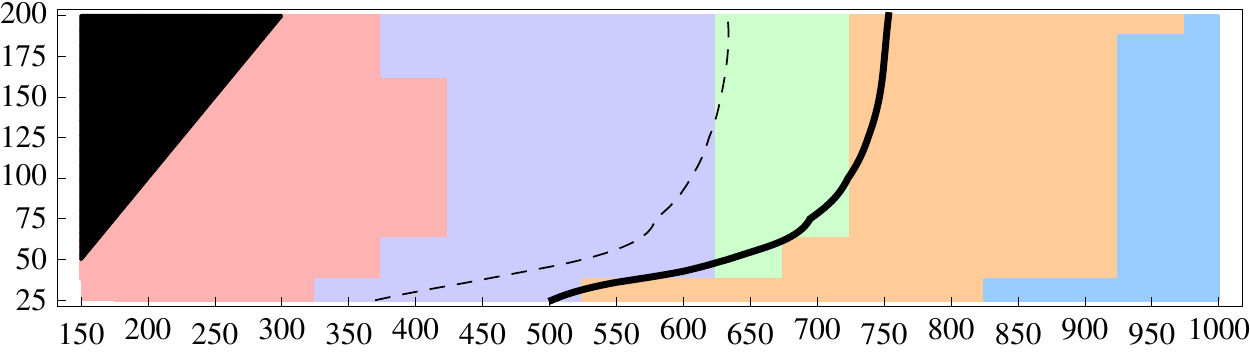}
\begin{picture}(0,0)(0,0)
\large
\put(-30,-8){$m_{\st}$}
\put(-490,70){$\delta$}
\put(-370,75){\bf R1}
\put(-270,75){\bf R2}
\put(-190,75){\bf R3}
\put(-125,75){\bf R4}
\put(-45,75){\bf R5}
\put(-428,100){\enwhiten{LEP}}
\normalsize
\end{picture}
\begin{minipage}[t]{7.0in}%
\caption{\label{fig:ExclusionLimits} Shown in the plane of $m_\st$ vs $\delta \equiv m_{\st}-m_{\cho^\pm}$ are the 95\% CL limits derived from the number of ``observed'' events as estimated in Table~\ref{tab:bkgds}.  The thick, solid (thin, dashed) line corresponds to the conservatively defined signal-only, i.e.~derived assuming expected background is zero as discussed in the text, 95\% exclusion limit for a 25\% (50\%) systematic uncertainty applied to the signal.   The five signal regions are denoted by R1, R2, etc., and indicate the most constraining signal region in the case of 25\% signal systematic uncertainty.  In the upper left corner of the plot, $m_\cho$ falls below 100 GeV, and would either have appeared at LEP or has a mass that is unphysical.}
\end{minipage}
\end{center}
\end{figure}
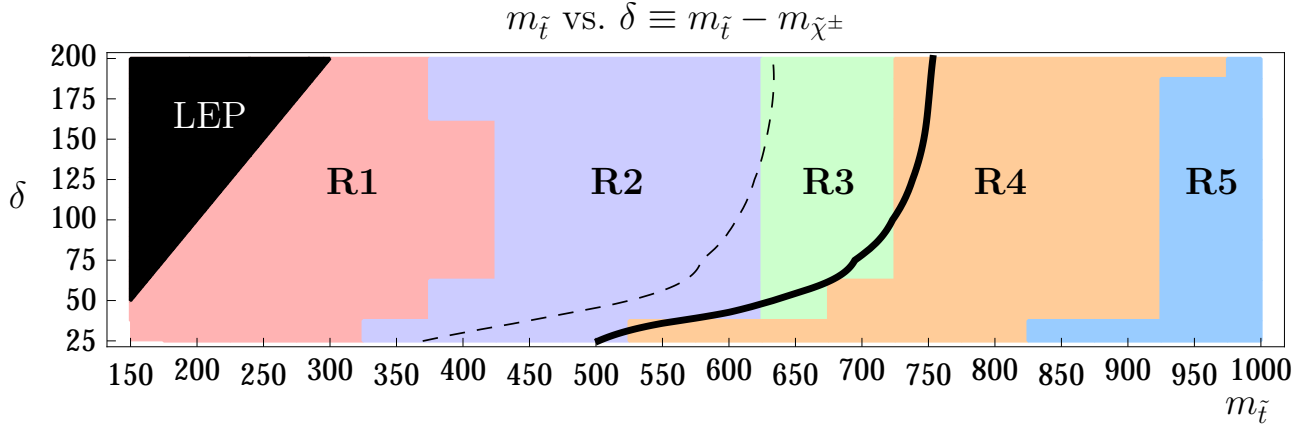 

As can be clearly seen in Table~\ref{tab:bkgds}, our signals over much of the interesting region are expected to be much larger, often by even an order of magnitude, than our backgrounds.  While one might not trust our background estimates, we can assume solely for the sake of deriving extremely conservative exclusion limits that our expected backgrounds are zero.  If our expected number of signal events proves too impossibly large to have only produced the number of observed events, then the signal can be excluded with some confidence.  For instance, if a study were performed and Region 2 were measured to have 6 events, whereas our signal predicts 50 events (as with our $m_\st=500$ GeV; $m_\cho=350$ benchmark), then it is extremely improbable that this signal would have such a large downward fluctuation to produce a mere 6 events.  This can be well excluded, certainly at a 95\% confidence level, even with fairly large systematic uncertainties on the signal.  This conservative exclusion method is only useful in placing robust limits.  In particular, in the event of a genuine signal, this method would be of no use for discovery.  

This conservative exclusion method of assuming the backgrounds are zero can be applied across the five signal regions to constrain our benchmark RPV model.  To simulate our signal, we use a grid of points in $m_\st \in [150,1000]$ GeV in 50 GeV steps vs $m_{\cho^\pm}=[m_\st-25$ GeV$,m_\st-200$ GeV$]$ in 25 GeV steps.  These signal samples were generated using {\tt MadGraph 5} and showered in {\tt Pythia8}.\footnote{As in \cite{Bai:2013xla}, we set {\tt SpaceShower::ptDampMatch=1} \cite{Corke:2010zj}, as this has been shown to more reliably mirror a matched sample for pair-produced stop events.  While a fully matched sample would be expected to reflect a realistic signature more accurately, the computational cost on such a high multiplicity signature is too great for that method to be viable in this phenomenological study.}  

In Figure~\ref{fig:ExclusionLimits}, the exclusion contours, assuming the data observed manifests as in Table \ref{tab:bkgds}, are shown as a thick, solid (thin, dashed) black line for signal systematic uncertainties of $25\% \; (50\%)$, where the signal uncertainty is assumed to be gaussian.  The distribution of the five signal regions shown is for the case with $25\%$ systematic signal uncertainty.   As can be clearly seen, even under the conservative assumptions used here, powerful exclusion on the signal region is achievable.  Of course, for the same number of observed events, exclusions would become stronger by treating any background estimation acquired from data-driven methods. 

\subsection{Signal discovery}
\label{sec:sigDisc}

\begin{figure}[t]
\begin{center}
\hspace{9mm}
\includegraphics[scale=0.58]{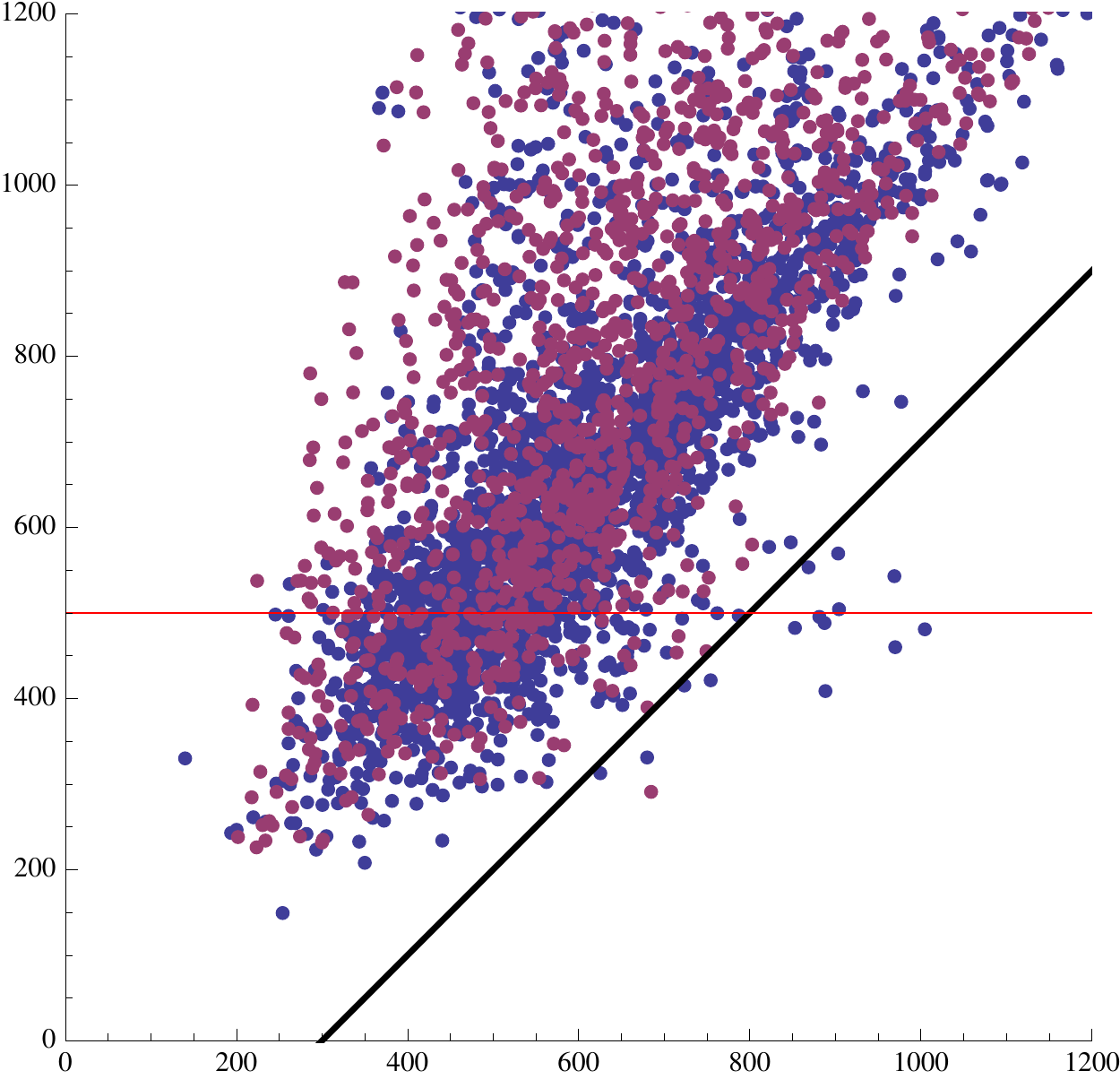}
\hspace{15mm}
\includegraphics[scale=0.58]{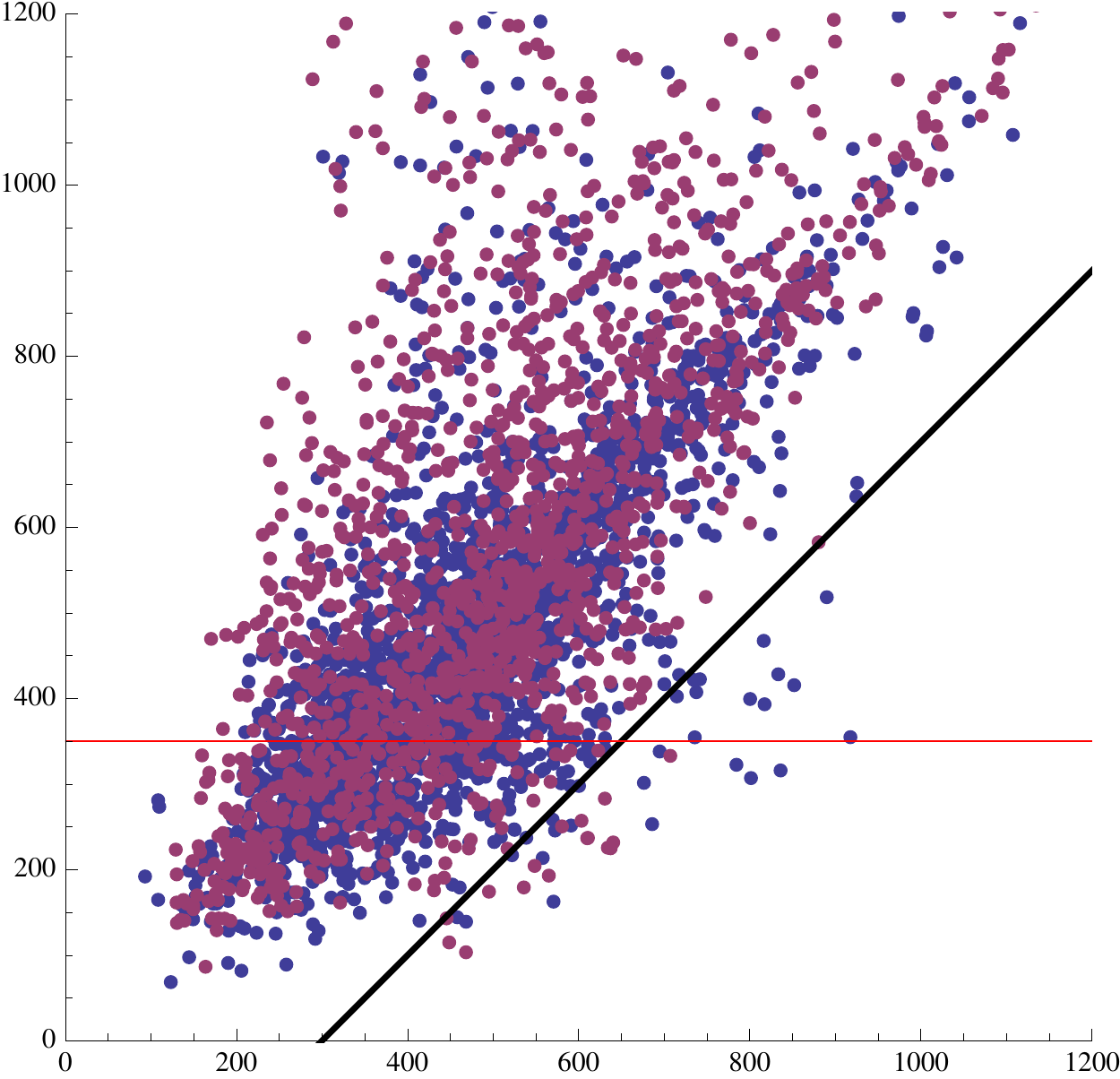}
\begin{picture}(0,0)(0,0)
\large
\put(-500,190){$m_{4j}$}
\put(-241,190){$m_{3j}$}
\put(-320,-14){$\sum_4 \abs{p_{T,j}}$}
\put(-60,-14){$\sum_3\abs{p_{T,j}}$}
\normalsize
\end{picture}
\begin{minipage}[t]{7.0in}%
\caption{\label{fig:mjvpt}  Scatter plot of all combinatoric possibilities of four (left) and three (right) jets constructed from the leading (up to) eight jets in an event passing the selection of Region 2.  Red points are 50 background events (in $b\bar bb\bar b+\{$jets$\}$).  Blue points are 50 signal events of $m_\st=500$ GeV and $m_{\cho^\pm}=350$ GeV (masses shown by horizontal red line).  The area to the right of the diagonal black line (offset by $\Delta_j=300$ GeV) indicates the signal region.  The signal region is, in both cases, populated dominantly by signal events, especially in the vicinity of the physical masses.}
\end{minipage}
\end{center}
\end{figure}
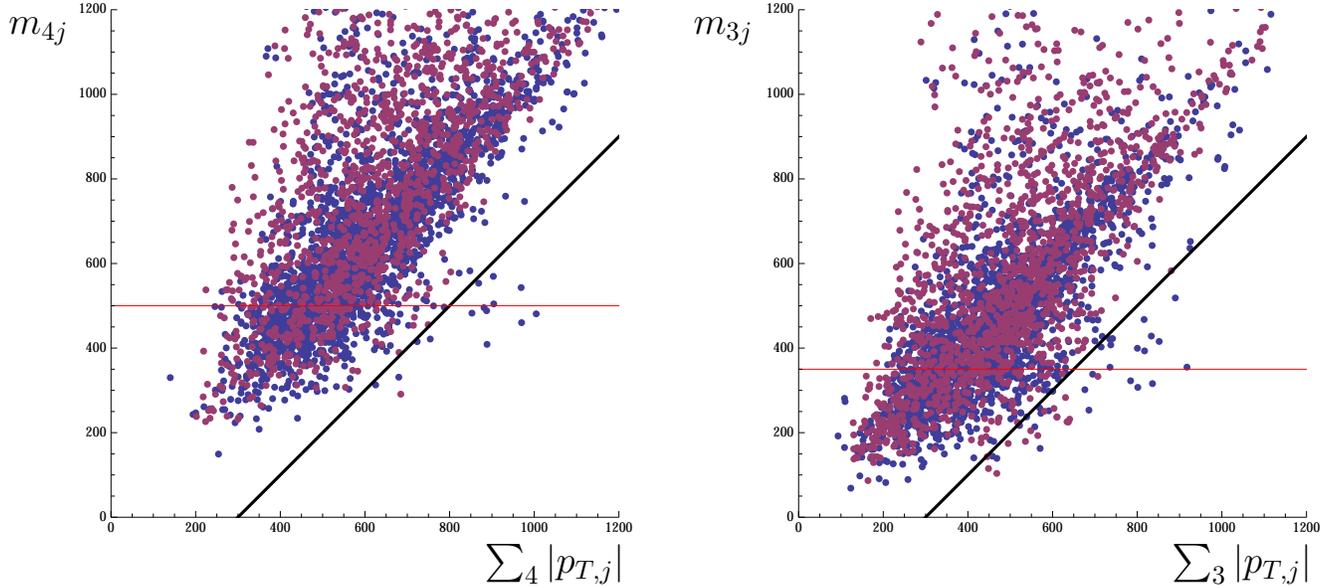 

Unfortunately, the method utilized in the previous section is only capable of setting limits and could never indicate an excess as being due to a genuine signal.  In the face of very uncertain backgrounds, how can one make a discovery?  Fortunately, these exotic events are, in principle, fully reconstructable.  If a larger than anticipated background is observed, then one can utilize the ``jet-ensemble'' technique of \cite{Aaltonen:2011sg,Chatrchyan:2011cj,Chatrchyan:2012uxa,Chatrchyan:2013gia} to search for resonant structures.

In these CDF and CMS searches for RPV gluinos, a scatter plot is formed by taking each event which passes the selection cuts, considering all possible trijet combinations, and plotting $\sum_i \abs{p_{T,j}}$ vs $M_{jjj}$.  As the trijet objects originating from a true resonance will occasionally be somewhat boosted, the decay products will be slightly collimated and appear with a higher $\sum_i \abs{p_{T,j}}$, while preserving an $M_{jjj}$ value close in mass to the resonance.  On the other hand, objects that do not all originate from a common resonance typically have $M_{jjj} \gtrsim \sum_i \abs{p_{T,j}}$.  Thus, by only considering events populating a region offset by $\sum_j \abs{p_{T,j}}-M_{jjj}>\Delta_j$, both the incorrect jet combinations in signal events, and all combinations in the typical background events are removed.  

We will adopt this strategy by taking the leading eight jets with $p_{T,j}>30$ GeV in each event (if fewer than eight jets are in an event passing the selection criteria, we take them all) and consider all possible combinations to form both trijet and quadrajet resonances (for eight jets, this means 56 and 70 combinations, respectively).  We will implement a very large offset of $\Delta_j=300$ GeV.   These distributions are presented in Figure~\ref{fig:mjvpt} for our $m_\st=500$ GeV and $m_{\cho^\pm}=350$ GeV benchmark in signal region 2.  Shown in blue are 50 events from this signal; in red are 50 events originating from a scaled up $b\bar bb\bar b+\{$light jets$\}$ background (with a $K$-factor of $\sim 30$ applied).  Even with only 50 events, the distribution of signal is markedly different from that of the background within the signal region (lower right region of each plot).  In models producing fewer events, the higher energy LHC run may be necessary to determine that an excess originates from a genuine signal, as opposed to a higher than expected background.

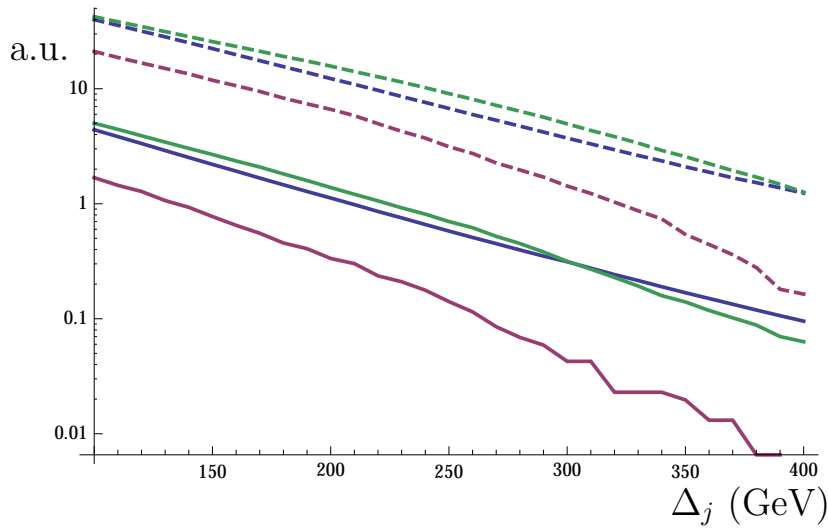
\begin{figure}[t]
\begin{center}
\hspace{9mm}
\includegraphics[scale=0.8]{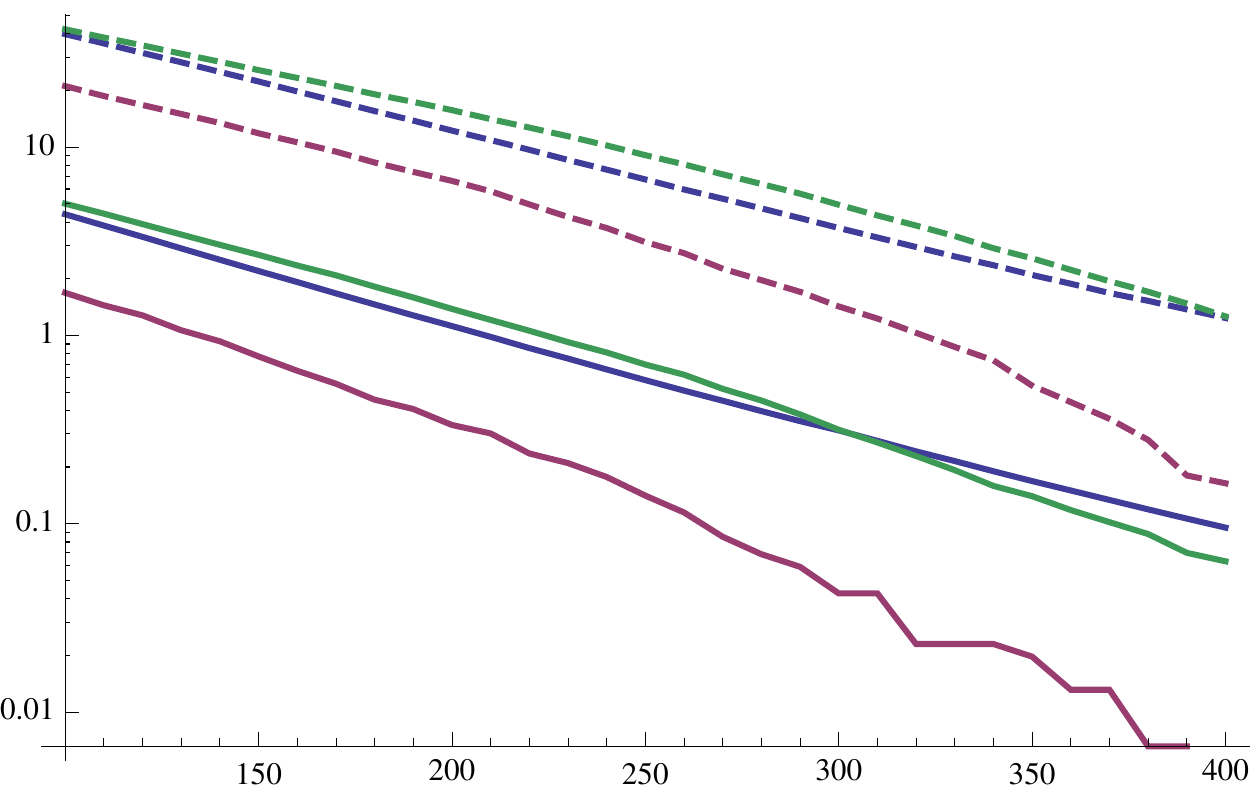}
\begin{picture}(0,0)(0,0)
\large
\put(-310,160){a.u.}
\put(-60,-12){$\Delta_{j}$ (GeV)}
\normalsize
\end{picture}
\begin{minipage}[t]{7.0in}%
\caption{\label{fig:mjwithcuts}  The fraction of combinations occupying the signal region, $\sum\abs{p_{T,j}}-m_j>\Delta_j$ (e.g., right of the black line in Figure~\ref{fig:mjvpt}) for four-jet (solid) and three-jet (dashed) combinations.  The curves are normalized so that the same number of events pass the selection cuts of Region 2 (all dashed lines are scaled up by a factor of $5$ to improve visibility).   Red curves are background events (in $b\bar bb\bar b+\{$jets$\}$).  Blue (green) curves are signal events with $m_\st=500$ GeV and $m_{\cho^\pm}=350$ GeV ($m_\st=300$ GeV and $m_{\cho^\pm}=200$ GeV).  Note that the background is significantly less for both signals (blue and green) and both combinations (3j and 4j), and this deviation grows more pronounced at larger $\Delta_j$. }
\end{minipage}
\end{center}
\end{figure}

In Figure \ref{fig:mjvpt}, a $\Delta_j$ cut of 300 GeV was used, but the results are not very sensitive to this precise choice.  A normalized fraction of combinations passing the $\Delta_j$ cut is shown in Figure~\ref{fig:mjwithcuts} for both four-jet (solid) and three-jet (dashed) reconstructions.  There, it can be seen that increasing $\Delta_j$ more harshly affect background than signal.  Even a mild choice of $\Delta_j$ (e.g., 100 GeV) can discriminate signal from background, but a larger value provides better discrimination if there are sufficiently many events.  This fact has no significant dependence on the spectra of signal masses.  

The kinematic features of the signal and background samples shown in Figure \ref{fig:mjvpt} should hold even if fully matched samples were generated.  As the signal distribution is due to the decay of slightly boosted resonances, this feature should be unaffected by matrix element / parton shower matching.  If {\tt Alpgen} was equipped to match the background samples with an MLM matching procedure, the matched events would still be a subset of the unmatched events used in this study.  Matching could enhance the number of background events populating the signal region only if there were an abnormally large matching efficiency ratio between background events that reach the signal region and events that do not.  As there is no reason to expect such a behavior, and nothing such as this appears in the QCD data of \cite{Chatrchyan:2013gia}, it should not happen to these high $b$-jet multiplicity samples.  

Of course, because the simulation done here is at leading order with unmatched jets, the uncertainties on the background are potentially large, especially the uncertainties in the kinematic distributions.  Data-driven control regions should be used to more accurately estimate the kinematic distributions of the backgrounds.  In an actual experimental study, one could utilize the data from lower $n_b$ channels, i.e. $n_b=0,1,2,3,4$ regions, to estimate the expected kinematic distributions of the jets in the $n_b=5$ sample.  In particular, the number of events expected to fall into the signal region at particular $m_{3j}$ and $m_{4j}$ values, should be predictable.  Importantly, the contamination from the top samples in each $n_b$ region needs to be correctly extrapolated.  The details of implementing this data-driven approach are better suited for the experimenters conducting the study. 

While not relevant for the particular model studied here, the other models producing high $b$-jet multiplicity signatures that were mentioned in Section~\ref{sec:NP} all contained $b\bar b$ resonances.  Resonances in $b\bar b$ are quite generic, and identifying them within an excess would be a key to understanding the signal, so we will briefly discuss how one could distinguish them from background.  This discussion is only intended to indicate that there are simple handles on the problem that can be employed to discover these scenarios in the event of an excess, and is not intended to be a thorough study of these $b\bar b$ resonance scenarios.

First, one could plot the dijet masses versus sum $p_T$ formed only from $b$-tagged jets.  This would help to reduce combinatoric inefficiency.  For the stealth gluino example, one would expect decay products to often be quite boosted.  However, in cases where the boost is not typically as significant (for instance, the higgsino production example), this method would be less effective.  As another option, one could, for each event, choose the two $b\bar b$ pairs with the greatest $\Delta=\abs{p_{T,b_1}}+\abs{p_{T,b_2}} - M_{bb}$, and plot $M_{bb,1}$ vs $M_{bb,2}$ to locate resonances.  This method would obviously be most effective when all resonance are of the same mass, although with enough events, it could suffice even if there are multiple $b\bar b$ resonances of different masses.

\section{Conclusions}
\label{sec:conclusions}

In this work, we have discussed the value and feasibility of searching for new physics within very high $b$-jet multiplicities.  We utilized signal regions that focused on the $H_T$-based trigger at CMS, and that contained five or more $b$-tagged jets.   With the conservative estimates used in this study, these channels are shown to have very low expected backgrounds, which should yield only a handful of events in the existing 20 fb$^{-1}$ data set at 8 TeV.  

To study the potential for uncovering new physics in this channel, we focused on a particular $R$-parity violating and minimal flavor violating supersymmetric scenario with a natural superpartner mass hierarchy containing only a right-handed stop and nearly degenerate higgsinos.  This model, with only a few parameters, receives no constraints from any existing experimental study.  Due to the large background uncertainties, we adopted an asymmetric approach to constrain and discover our signal through different methods.  With our study, we illustrated that powerful constraints are feasible, even under extremely conservative assumptions about the size of the background.  We showed that stops up to 750 GeV could realistically be excluded by a study using the current data set. 

The asymmetric method, cut-and-count for exclusion and resonant reconstruction for discovery, can be applied to other scenarios with very uncertain backgrounds as well.  For instance, the backgrounds to another potential signature of RPV stops $\st \to t \cho \to t (jjj)$ has extremely large uncertainties, however, the $t\bar t +6$ jet background is small enough that meaningful limits could be placed (as can be inferred from \cite{ATLAS:2012ceu}).  Resonant reconstruction on the same sample could distinguish new physics from a large background.

While one of the most motivated RPV SUSY scenarios can give rise to this unconstrained signature, we discussed a handful of other models that can also yield high $b$-jet multiplicities.  These specific examples included stealth SUSY, $b'$ models, and extended Higgs sectors.  The search strategy employed in this work is generic enough that other models of new physics, especially those that couple to standard model particles with a third-generation dominant structure, could be constrained or discovered through an experimental study of this nature.  Models such as these can slip through the current LHC analyses, meaning this signature is an outstanding gap in LHC coverage at both ATLAS and CMS.  

Even in the absence of new physics, a measurement of high $b$-jet multiplicities at the LHC is important, as it can help to further our understanding of QCD, while simultaneously providing important constraints on models that have not yet been imagined.  When the LHC begins collecting data at a higher energy, the results from 8 TeV will be invaluable for background estimation.  For these reasons, such an experimental study should use general cuts and not tailor itself specifically to any one signal in order to maximize its value in the future.

\section*{Acknowledgments}
\noindent
We thank J.~Chou, B.~Dahmes, J.~Dolen, D.~Ferencek, R.~Franceschini, E.~Halkiadakis, V.~Hirschi, Y.~Kats, A.~Lath, M.~Strassler, A.~Thalapillil, and S.~Thomas for useful conversations.  We thank N.~Craig and S.~Thomas for reading the draft and providing useful feedback.  This research was supported by DOE grants DOE-SC0010008, DOE-ARRA-SC0003883, and DOE-DE-SC0007897.

\small{
\bibliography{Swarm}}

\providecommand{\href}[2]{#2}\begingroup\raggedright\begin{thebibliography}{10}

\bibitem{Chatrchyan:2012ufa}
{ CMS} collaboration, S.~Chatrchyan {\em et~al.}, ``{Observation of a new boson
  at a mass of 125~GeV with the CMS experiment at the LHC},''
  \href{http://dx.doi.org/10.1016/j.physletb.2012.08.021}{{\em Phys.Lett.}
  {\bfseries B716} (2012) 30},
\href{http://arxiv.org/abs/1207.7235}{{\ttfamily arXiv:1207.7235 [hep-ex]}}.

\bibitem{Aad:2012tfa}
{ ATLAS} collaboration, G.~Aad {\em et~al.}, ``{Observation of a new particle
  in the search for the Standard Model Higgs boson with the ATLAS detector at
  the LHC},'' \href{http://dx.doi.org/10.1016/j.physletb.2012.08.020}{{\em
  Phys.Lett.} {\bfseries B716} (2012) 1},
\href{http://arxiv.org/abs/1207.7214}{{\ttfamily arXiv:1207.7214 [hep-ex]}}.

\bibitem{Craig:2013cxa}
N.~Craig, ``{The State of Supersymmetry after Run I of the LHC},''
\href{http://arxiv.org/abs/1309.0528}{{\ttfamily arXiv:1309.0528 [hep-ph]}}.

\bibitem{Feng:2013pwa}
J.~L. Feng, ``{Naturalness and the Status of Supersymmetry},''
  \href{http://dx.doi.org/10.1146/annurev-nucl-102010-130447}{{\em
  Ann.Rev.Nucl.Part.Sci.} {\bfseries 63} (2013) 351--382},
\href{http://arxiv.org/abs/1302.6587}{{\ttfamily arXiv:1302.6587 [hep-ph]}}.

\bibitem{Randall:2008rw}
L.~Randall and D.~Tucker-Smith, ``{Dijet Searches for Supersymmetry at the
  LHC},'' \href{http://dx.doi.org/10.1103/PhysRevLett.101.221803}{{\em
  Phys.Rev.Lett.} {\bfseries 101} (2008) 221803},
\href{http://arxiv.org/abs/0806.1049}{{\ttfamily arXiv:0806.1049 [hep-ph]}}.

\bibitem{Rogan:2010kb}
C.~Rogan, ``{Kinematical variables towards new dynamics at the LHC},''
\href{http://arxiv.org/abs/1006.2727}{{\ttfamily arXiv:1006.2727 [hep-ph]}}.

\bibitem{CMS:2013qua}
{ CMS} collaboration, ``{A search for new physics in events with high jet and
  b-tagged jet multiplicities and one lepton},''
  \href{http://cds.cern.ch/record/1632190}{CMS-PAS-SUS-12-015} (2013).

\bibitem{Craig:2012vn}
N.~Craig and S.~Thomas, ``{Exclusive Signals of an Extended Higgs Sector},''
  \href{http://dx.doi.org/10.1007/JHEP11(2012)083}{{\em JHEP} {\bfseries 1211}
  (2012) 083},
\href{http://arxiv.org/abs/1207.4835}{{\ttfamily arXiv:1207.4835 [hep-ph]}}.

\bibitem{Rubin:2010xp}
M.~Rubin, G.~P. Salam, and S.~Sapeta, ``{Giant QCD K-factors beyond NLO},''
  \href{http://dx.doi.org/10.1007/JHEP09(2010)084}{{\em JHEP} {\bfseries 1009}
  (2010) 084},
\href{http://arxiv.org/abs/1006.2144}{{\ttfamily arXiv:1006.2144 [hep-ph]}}.

\bibitem{D'Ambrosio:2002ex}
G.~D'Ambrosio, G.~Giudice, G.~Isidori, and A.~Strumia, ``{Minimal flavor
  violation: An Effective field theory approach},''
  \href{http://dx.doi.org/10.1016/S0550-3213(02)00836-2}{{\em Nucl.Phys.}
  {\bfseries B645} (2002) 155--187},
\href{http://arxiv.org/abs/hep-ph/0207036}{{\ttfamily arXiv:hep-ph/0207036
  [hep-ph]}}.

\bibitem{Agashe:2005hk}
K.~Agashe, M.~Papucci, G.~Perez, and D.~Pirjol, ``{Next to minimal flavor
  violation},''
\href{http://arxiv.org/abs/hep-ph/0509117}{{\ttfamily arXiv:hep-ph/0509117
  [hep-ph]}}.

\bibitem{Nomura:2007ap}
Y.~Nomura, M.~Papucci, and D.~Stolarski, ``{Flavorful supersymmetry},''
  \href{http://dx.doi.org/10.1103/PhysRevD.77.075006}{{\em Phys.Rev.}
  {\bfseries D77} (2008) 075006},
\href{http://arxiv.org/abs/0712.2074}{{\ttfamily arXiv:0712.2074 [hep-ph]}}.

\bibitem{Evans:2013kxa}
J.~A. Evans and D.~Shih, ``{Surveying Extended GMSB Models with mh=125 GeV},''
  \href{http://dx.doi.org/10.1007/JHEP08(2013)093}{{\em JHEP} {\bfseries 1308}
  (2013) 093},
\href{http://arxiv.org/abs/1303.0228}{{\ttfamily arXiv:1303.0228 [hep-ph]}}.

\bibitem{Papucci:2011wy}
M.~Papucci, J.~T. Ruderman, and A.~Weiler, ``{Natural SUSY Endures},''
  \href{http://dx.doi.org/10.1007/JHEP09(2012)035}{{\em JHEP} {\bfseries 1209}
  (2012) 035},
\href{http://arxiv.org/abs/1110.6926}{{\ttfamily arXiv:1110.6926 [hep-ph]}}.

\bibitem{Fox:2002bu}
P.~J. Fox, A.~E. Nelson, and N.~Weiner, ``{Dirac gaugino masses and supersoft
  supersymmetry breaking},'' {\em JHEP} {\bfseries 0208} (2002) 035,
\href{http://arxiv.org/abs/hep-ph/0206096}{{\ttfamily arXiv:hep-ph/0206096
  [hep-ph]}}.

\bibitem{Kribs:2012gx}
G.~D. Kribs and A.~Martin, ``{Supersoft Supersymmetry is Super-Safe},''
  \href{http://dx.doi.org/10.1103/PhysRevD.85.115014}{{\em Phys.Rev.}
  {\bfseries D85} (2012) 115014},
\href{http://arxiv.org/abs/1203.4821}{{\ttfamily arXiv:1203.4821 [hep-ph]}}.

\bibitem{Nikolidakis:2007fc}
E.~Nikolidakis and C.~Smith, ``{Minimal Flavor Violation, Seesaw, and
  R-parity},'' \href{http://dx.doi.org/10.1103/PhysRevD.77.015021}{{\em
  Phys.Rev.} {\bfseries D77} (2008) 015021},
\href{http://arxiv.org/abs/0710.3129}{{\ttfamily arXiv:0710.3129 [hep-ph]}}.

\bibitem{Csaki:2011ge}
C.~Csaki, Y.~Grossman, and B.~Heidenreich, ``{MFV SUSY: A Natural Theory for
  R-Parity Violation},''
  \href{http://dx.doi.org/10.1103/PhysRevD.85.095009}{{\em Phys.Rev.}
  {\bfseries D85} (2012) 095009},
\href{http://arxiv.org/abs/1111.1239}{{\ttfamily arXiv:1111.1239 [hep-ph]}}.

\bibitem{Krnjaic:2012aj}
G.~Krnjaic and D.~Stolarski, ``{Gauging the Way to MFV},''
  \href{http://dx.doi.org/10.1007/JHEP04(2013)064}{{\em JHEP} {\bfseries 1304}
  (2013) 064},
\href{http://arxiv.org/abs/1212.4860}{{\ttfamily arXiv:1212.4860 [hep-ph]}}.

\bibitem{Franceschini:2013ne}
R.~Franceschini and R.~Mohapatra, ``{New Patterns of Natural R-Parity Violation
  with Supersymmetric Gauged Flavor},''
  \href{http://dx.doi.org/10.1007/JHEP04(2013)098}{{\em JHEP} {\bfseries 1304}
  (2013) 098},
\href{http://arxiv.org/abs/1301.3637}{{\ttfamily arXiv:1301.3637 [hep-ph]}}.

\bibitem{Csaki:2013we}
C.~Csaki and B.~Heidenreich, ``{A Complete Model for R-parity Violation},''
  \href{http://dx.doi.org/10.1103/PhysRevD.88.055023}{{\em Phys.Rev.}
  {\bfseries D88} (2013) 055023},
\href{http://arxiv.org/abs/1302.0004}{{\ttfamily arXiv:1302.0004 [hep-ph]}}.

\bibitem{Franceschini:2012za}
R.~Franceschini and R.~Torre, ``{RPV stops bump off the background},''
  \href{http://dx.doi.org/10.1140/epjc/s10052-013-2422-x}{{\em Eur.Phys.J.}
  {\bfseries C73} (2013) 2422},
\href{http://arxiv.org/abs/1212.3622}{{\ttfamily arXiv:1212.3622 [hep-ph]}}.

\bibitem{Bai:2013xla}
Y.~Bai, A.~Katz, and B.~Tweedie, ``{Pulling Out All the Stops: Searching for
  RPV SUSY with Stop-Jets},''
\href{http://arxiv.org/abs/1309.6631}{{\ttfamily arXiv:1309.6631 [hep-ph]}}.

\bibitem{Evans:2012bf}
J.~A. Evans and Y.~Kats, ``{LHC Coverage of RPV MSSM with Light Stops},''
  \href{http://dx.doi.org/10.1007/JHEP04(2013)028}{{\em JHEP} {\bfseries 1304}
  (2013) 028},
\href{http://arxiv.org/abs/1209.0764}{{\ttfamily arXiv:1209.0764 [hep-ph]}}.

\bibitem{Evans:2013uwa}
J.~A. Evans and Y.~Kats, ``{LHC searches examined via the RPV MSSM},''
\href{http://arxiv.org/abs/1311.0890}{{\ttfamily arXiv:1311.0890 [hep-ph]}}.

\bibitem{Strassler:2006im}
M.~J. Strassler and K.~M. Zurek, ``{Echoes of a hidden valley at hadron
  colliders},'' \href{http://dx.doi.org/10.1016/j.physletb.2007.06.055}{{\em
  Phys.Lett.} {\bfseries B651} (2007) 374--379},
\href{http://arxiv.org/abs/hep-ph/0604261}{{\ttfamily arXiv:hep-ph/0604261
  [hep-ph]}}.

\bibitem{Strassler:2008fv}
M.~J. Strassler, ``{On the Phenomenology of Hidden Valleys with Heavy
  Flavor},''
\href{http://arxiv.org/abs/0806.2385}{{\ttfamily arXiv:0806.2385 [hep-ph]}}.

\bibitem{Fan:2011yu}
J.~Fan, M.~Reece, and J.~T. Ruderman, ``{Stealth Supersymmetry},''
  \href{http://dx.doi.org/10.1007/JHEP11(2011)012}{{\em JHEP} {\bfseries 1111}
  (2011) 012},
\href{http://arxiv.org/abs/1105.5135}{{\ttfamily arXiv:1105.5135 [hep-ph]}}.

\bibitem{Fan:2012jf}
J.~Fan, M.~Reece, and J.~T. Ruderman, ``{A Stealth Supersymmetry Sampler},''
  \href{http://dx.doi.org/10.1007/JHEP07(2012)196}{{\em JHEP} {\bfseries 1207}
  (2012) 196},
\href{http://arxiv.org/abs/1201.4875}{{\ttfamily arXiv:1201.4875 [hep-ph]}}.

\bibitem{Evans:2013jna}
J.~A. Evans, Y.~Kats, D.~Shih, and M.~J. Strassler, ``{Toward Full LHC Coverage
  of Natural Supersymmetry},''
\href{http://arxiv.org/abs/1310.5758}{{\ttfamily arXiv:1310.5758 [hep-ph]}}.

\bibitem{Craig:2013hca}
N.~Craig, J.~Galloway, and S.~Thomas, ``{Searching for Signs of the Second
  Higgs Doublet},''
\href{http://arxiv.org/abs/1305.2424}{{\ttfamily arXiv:1305.2424 [hep-ph]}}.

\bibitem{Aguilar-Saavedra:2013qpa}
J.~Aguilar-Saavedra, R.~Benbrik, S.~Heinemeyer, and M.~Perez-Victoria, ``{A
  handbook of vector-like quarks: mixing and single production},''
  \href{http://dx.doi.org/10.1103/PhysRevD.88.094010}{{\em Phys.Rev.}
  {\bfseries D88} (2013) 094010},
\href{http://arxiv.org/abs/1306.0572}{{\ttfamily arXiv:1306.0572 [hep-ph]}}.

\bibitem{Azatov:2012rj}
A.~Azatov, O.~Bondu, A.~Falkowski, M.~Felcini, S.~Gascon-Shotkin, {\em et~al.},
  ``{Higgs boson production via vector-like top-partner decays: Diphoton or
  multilepton plus multijets channels at the LHC},''
  \href{http://dx.doi.org/10.1103/PhysRevD.85.115022}{{\em Phys.Rev.}
  {\bfseries D85} (2012) 115022},
\href{http://arxiv.org/abs/1204.0455}{{\ttfamily arXiv:1204.0455 [hep-ph]}}.

\bibitem{Chatrchyan:2012jua}
{ CMS} collaboration, S.~Chatrchyan {\em et~al.}, ``{Identification of b-quark
  jets with the CMS experiment},''
  \href{http://dx.doi.org/10.1088/1748-0221/8/04/P04013}{{\em JINST} {\bfseries
  8} (2013) P04013},
\href{http://arxiv.org/abs/1211.4462}{{\ttfamily arXiv:1211.4462 [hep-ex]}}.

\bibitem{ATLAS:2013tma}
{ ATLAS} collaboration, ``{Search for strongly produced superpartners in final
  states with two same sign leptons with the ATLAS detector using 21~fb$^{-1}$
  of proton-proton collisions at $\sqrt s = 8$~TeV},''
  \href{http://cds.cern.ch/record/1522430}{ATLAS-CONF-2013-007} (2013).

\bibitem{BDahmes}
 B.~Dahmes, private communication (2013).

\bibitem{CMS:2013vui}
{ CMS} collaboration, ``{Measurement of the cross section ratio $\sigma(t\bar
  tb\bar b)/(t\bar tjj)$ in pp Collisions at $\sqrt{s}=8$ TeV},''
  \href{http://cds.cern.ch/record/1605842}{CMS-PAS-TOP-13-010} (2013).

\bibitem{Mangano:2002ea}
M.~L. Mangano, M.~Moretti, F.~Piccinini, R.~Pittau, and A.~D. Polosa,
  ``{ALPGEN, a generator for hard multiparton processes in hadronic
  collisions},'' \href{http://dx.doi.org/10.1088/1126-6708/2003/07/001}{{\em
  JHEP} {\bfseries 0307} (2003) 001},
\href{http://arxiv.org/abs/hep-ph/0206293}{{\ttfamily arXiv:hep-ph/0206293
  [hep-ph]}}.

\bibitem{Sjostrand:2007gs}
T.~Sjostrand, S.~Mrenna, and P.~Z. Skands, ``{A Brief Introduction to PYTHIA
  8.1},'' \href{http://dx.doi.org/10.1016/j.cpc.2008.01.036}{{\em
  Comput.Phys.Commun.} {\bfseries 178} (2008) 852},
\href{http://arxiv.org/abs/0710.3820}{{\ttfamily arXiv:0710.3820 [hep-ph]}}.

\bibitem{Alwall:2011uj}
J.~Alwall, M.~Herquet, F.~Maltoni, O.~Mattelaer, and T.~Stelzer, ``{MadGraph 5
  : Going Beyond},'' \href{http://dx.doi.org/10.1007/JHEP06(2011)128}{{\em
  JHEP} {\bfseries 1106} (2011) 128},
\href{http://arxiv.org/abs/1106.0522}{{\ttfamily arXiv:1106.0522 [hep-ph]}}.

\bibitem{CMS:2012nca}
{ CMS} collaboration, ``{Search for Multijet Resonances in the 8-jet Final
  State},'' \href{http://cds.cern.ch/record/1482131}{CMS-PAS-EXO-11-075}
  (2012).

\bibitem{Corke:2010zj}
R.~Corke and T.~Sjostrand, ``{Improved Parton Showers at Large Transverse
  Momenta},'' \href{http://dx.doi.org/10.1140/epjc/s10052-010-1409-0}{{\em
  Eur.Phys.J.} {\bfseries C69} (2010) 1},
\href{http://arxiv.org/abs/1003.2384}{{\ttfamily arXiv:1003.2384 [hep-ph]}}.

\bibitem{Aaltonen:2011sg}
{ CDF} collaboration, T.~Aaltonen {\em et~al.}, ``{First Search for Multijet
  Resonances in $\sqrt{s} = 1.96$ TeV $ p\bar{p}$ Collisions},''
  \href{http://dx.doi.org/10.1103/PhysRevLett.107.042001}{{\em Phys.Rev.Lett.}
  {\bfseries 107} (2011) 042001},
\href{http://arxiv.org/abs/1105.2815}{{\ttfamily arXiv:1105.2815 [hep-ex]}}.

\bibitem{Chatrchyan:2011cj}
{ CMS} collaboration, S.~Chatrchyan {\em et~al.}, ``{Search for Three-Jet
  Resonances in $pp$ Collisions at $\sqrt{s}=7$ TeV},''
  \href{http://dx.doi.org/10.1103/PhysRevLett.107.101801}{{\em Phys.Rev.Lett.}
  {\bfseries 107} (2011) 101801},
\href{http://arxiv.org/abs/1107.3084}{{\ttfamily arXiv:1107.3084 [hep-ex]}}.

\bibitem{Chatrchyan:2012uxa}
{ CMS} collaboration, S.~Chatrchyan {\em et~al.}, ``{Search for three-jet
  resonances in $pp$ collisions at $\sqrt{s}=7$ TeV},''
  \href{http://dx.doi.org/10.1016/j.physletb.2012.10.048}{{\em Phys.Lett.}
  {\bfseries B718} (2012) 329--347},
\href{http://arxiv.org/abs/1208.2931}{{\ttfamily arXiv:1208.2931 [hep-ex]}}.

\bibitem{Chatrchyan:2013gia}
{ CMS} collaboration, S.~Chatrchyan {\em et~al.}, ``{Searches for light- and
  heavy-flavour three-jet resonances in $pp$ collisions at $\sqrt{s} = 8$
  TeV},''
\href{http://arxiv.org/abs/1311.1799}{{\ttfamily arXiv:1311.1799 [hep-ex]}}.

\bibitem{ATLAS:2012ceu}
{ ATLAS} collaboration, ``{Measurement of the jet multiplicity in top anti-top
  final states produced in 7 TeV proton-proton collisions with the ATLAS
  detector},'' \href{http://cds.cern.ch/record/1493494}{ATLAS-CONF-2012-155}
  (2012).

\end{thebibliography}\endgroup

\end{document}